\documentclass[journal]{IEEEtran}

\IEEEoverridecommandlockouts                              

\usepackage{changes}
\usepackage{authblk}
\usepackage{amsfonts}
\usepackage{amsmath,amssymb}       
\usepackage{hyperref,cite}
\usepackage{cleveref}
\usepackage{color}
\usepackage{algorithm}
\usepackage{algorithmic}
\let\labelindent\relax
\usepackage{enumitem}
\usepackage{float}
\usepackage{graphicx}
\usepackage{booktabs}
\usepackage{threeparttable}
\usepackage{hhline}
\newtheorem{theorem}{Theorem}

\newtheorem{remark}{Remark}
\newtheorem{lemma}{Lemma}
\DeclareMathOperator*{\argmin}{arg\,min} 
\newcommand{\diag}{\mathop{\rm diag}\nolimits}

\newcommand{\ba}[1]{\begin{array}{#1}}
\newcommand{\ea}{\end{array}}

\begin{document}
\title{\LARGE \bf
An Execution-time-certified Riccati-based IPM Algorithm for RTI-based Input-constrained NMPC}
\author{Liang Wu$^{1}$, Krystian Ganko$^{1}$, Shimin Wang$^{1}$, Richard D. Braatz$^{1}$, Fellow, IEEE%
\thanks{$^{1}$Massachusetts Institute of Technology, Cambridge, MA 02139, USA, {\tt\small \{liangwu,kkganko,bellewsm,braatz\}@mit.edu}\\
}
}

\maketitle

\begin{abstract}
Establishing an execution time certificate in deploying model predictive control (MPC) is a pressing and challenging requirement. As nonlinear MPC (NMPC) results in nonlinear programs, differing from quadratic programs encountered in linear MPC, deriving an execution time certificate for NMPC seems an impossible task. Our prior work \cite{wu2023direct} introduced an input-constrained MPC algorithm with the exact and only \textit{dimension-dependent} (\textit{data-independent}) number of floating-point operations ([flops]). This paper extends it to input-constrained NMPC problems via the real-time iteration (RTI) scheme, which results in \textit{data-varying} (but \textit{dimension-invariant}) input-constrained MPC problems. Therefore, applying our previous algorithm can certify the execution time based on the assumption that processors perform fixed [flops] in constant time. As the RTI-based scheme generally results in MPC with a long prediction horizon, this paper employs the efficient factorized Riccati recursion, whose computational cost scales linearly with the prediction horizon, to solve the Newton system at each iteration. The execution-time certified capability of the algorithm is theoretically and numerically validated through a case study involving nonlinear control of the chaotic Lorenz system.
\end{abstract}


\section{Introduction}
Model predictive control (MPC) is a model-based optimal control technique, which at each sampling time, handles the current feedback state information by solving an online constrained optimization formulated from a dynamical prediction model and user-specified constraints and objectives. 

Hence, a question arises---can the adopted MPC algorithm finish the optimization task handling the current feedback state information before the next feedback state information arrives? An execution-time certificate that guarantees that the MPC returns a control action before the next sampling time, is a necessity when deploying MPC in production environments.
In addition, theoretical execution time analysis can guide the trade-off of MPC settings, such as sampling time, adopted model complexity, and prediction horizon length, instead of using extensive simulations such as with the heavy calibration work of embedded MPC \cite{forgione2020efficient}.

This execution-time certificate requirement has garnered increasing attention within recent years and is still an active research area \cite{richter2011computational, bemporad2012simple, giselsson2012execution, cimini2017exact,arnstrom2021unifying}. All these works analyze the worst-case iteration bound of their proposed algorithms to derive a worst-case execution time based on the assumption that \textit{the adopted computation platform performs a fixed number of floating-point operations ([flops]) in constant time}. That is,
$$
\textrm{execution time} = \frac{\textrm{total [flops] required by the algorithm}}{\textrm{average [flops] processed per second}}~[s].
$$
This article also follows this assumption to certify the execution time of input-constrained nonlinear MPC (NMPC) problems.

Linear MPC is formulated using a linear process model, which leads to solving a convex quadratic program (QP) that can be efficiently solved using methods such as interior-point methods (IPM) \cite{wang2009fast}, active set methods \cite{ferreau2014qpoases}, and first-order methods\cite{stellato2020osqp, wu2023simple, wu2023construction}. NMPC instead adopts a nonlinear model, which results in a nonlinear program (NLP). NLPs not only have a higher computational burden but are also nearly impossible to derive the worst-case of required [flops] for their solution. For this reason, previous works \cite{richter2011computational, bemporad2012simple, giselsson2012execution, cimini2017exact, arnstrom2021unifying} on executive-time certificates forcused on linear MPC problems.
A straightforward idea to develop an execution-time certified NMPC algorithm is to apply these previous algorithms \cite{richter2011computational, bemporad2012simple, giselsson2012execution, cimini2017exact, arnstrom2021unifying} to some nonlinear-to-linear transformation techniques in the MPC field, such as successive online linearization or real-time iteration (RTI) \cite{gros2020linear}.

However, this idea is impractical because online linearized MPC problems (online linearization or RTI scheme-based MPC) have time-changing problem data, e.g., the time-changing Hessian matrix. In 
\cite{richter2011computational, bemporad2012simple, giselsson2012execution}, their derived computation complexity analysis is \textit{data-dependent}, namely depending on the data of the resulting optimization problem. In \cite{cimini2017exact, arnstrom2021unifying}, their worst-case computation complexity certification relies on the
complicated and computation-heavy (thus offline), which also limits their use in online linearized MPC problems. Thus, to the best of the authors' knowledge, no work extends these algorithms\cite{richter2011computational, bemporad2012simple, giselsson2012execution, cimini2017exact, arnstrom2021unifying} to NMPC problems.

Our recent work \cite{wu2023direct} for the first time 
proposes a box-constrained QP (Box-QP) algorithm with an exact and only \textit{dimension-dependent} (\textit{data-independent}) number of iterations $\mathcal{N}=\!\left\lceil\frac{\log\frac{2n}{\epsilon}}{-2\log\frac{\sqrt{2n}}{\sqrt{2n}+\sqrt{2}-1}}\right\rceil\! + 1,$
where $n$ denotes the problem dimension and $\epsilon$ denotes the constant (such as $1\times 10^{-6}$) stopping criterion. Therefore, our algorithm is ideally suited for online linearized-based NMPC schemes to certify the execution time.

Unlike the successive online linearization scheme, which only linearizes at the current point, the RTI scheme linearizes the whole previous trajectory and results in a linear time-varying model along the prediction horizon, thus providing a better nonlinear model approximation.
More importantly, the high-approximation RTI scheme does not increase the computational time thanks to the \textit{Preparation-Feedback Split}, in which the computation time is only marginally larger than linear MPC. Easy-to-understand tutorials with a detailed proof of nominal stability are available \cite{gros2020linear,diehl2001real,
diehl2005nominal}. 

Therefore, this paper adopts the RTI scheme for nonlinear input-constrained MPC problems and then applies our previous algorithm \cite{wu2023direct} can certify the execution time by analyzing the total required [flops]. Our previous work \cite{wu2024time} also employed the data-driven Koopman operator, which lifts the nonlinear system to a higher but linear system \cite{korda2018linear} for nonlinear input-constrained MPC problems to certify the execution time by using our previous algorithm \cite{wu2023direct}.
Compared to the data-driven Koopman operator, the advantage of the RTI-based scheme is the ability to use existing first-principles continuous-time models from physical laws, such as found in chemical engineering and robotics. As such, this work is a complementary approach to applications where a first-principles continuous-time model is available.

\subsection{Contributions}
This work, for the first time, develops an execution-time-certified algorithm for RTI-based input-constrained NMPC problems. 
The RTI-based scheme performs well, with a small sampling time and a long time horizon \cite{gros2020linear}, which results in MPC problems with a long prediction horizon. Then we employ an efficient factorized Riccati recursion characterized by a computational cost that scales linearly with the prediction horizon, facilitating the solution of the Newton system at each iteration. 
Additionally, as a byproduct, the utilized factorized Riccati recursion eliminates the need for computing the Hessian matrix, thereby reducing both the computational time and the implementation complexity of the preparation phase.

\subsection{Notation}
$\mathbb{R}^n$ denotes the space of $n$-dimensional real vectors, $\mathbb{R}^n_{++}$ is the set of all positive vectors of $\mathbb{R}^n$, and $\mathbb{N}_+$ is the set of positive integers. For a vector $z\in\mathbb{R}^n$, its Euclidean norm is $\|z\|=\sqrt{z_1^2+z_2^2+\cdots+z_n^2}$, $\|z\|_1=\sum_{i=1}^{n}|z_i|$, $\diag(z):\mathbb{R}^n\rightarrow\mathbb{R}^{n\times n}$ maps an vector $z$ to its corresponding diagonal matrix, and $z^2 = (z_1^2,z_2^2,\cdots{},z_n^2)^\top$. A function is defined as $\| x \|_{Q}^{2} = x^{\top} Q x$. Given two arbitrary vectors $z,y \in\mathbb{R}^n_{++}$, their Hadamard product is $zy = (z_1y_1,z_2y_2,\cdots{},z_ny_n)^\top$, $\big(\frac{z}{y}\big)=\big(\frac{z_1}{y_1},\frac{z_2}{y_2},\cdots{},\frac{z_n}{y_n}
\big)^{\!\top}$,  $\sqrt{z}=\left(\sqrt{z_1},\sqrt{z_2},\cdots{},\sqrt{z_n}\right)^{\!\top}$. The vector of all ones is denoted by $e=(1,\cdots{},1)^\top$. $\left\lceil x\right\rceil$ maps $x$ to the least integer greater than or equal to $x$. For $z,y\in\mathbb{R}^n$, let $\mathrm{col}(z,y)=[z^{\top},y^{\top}]^{\top}$.

\section{RTI-based input-constrained NMPC}
In this article, we consider the tracking input-constrained NMPC problem $\text{NLP}\equiv \text{NLP}(\hat{x}_t,\mathbf{x}_t^{\text{ref}},\mathbf{u}_t^{\text{ref}})$: 

\begin{equation}\label{eqn_RTI_NMPC}
    \begin{aligned}
        \text{NLP} &\triangleq \argmin  \tfrac{1}{2}\|x_{t,N}-x_{t,N}^{\text{ref}}\|_{W_N}^2 \\
        & +\sum_{k=0}^{N-1}\tfrac{1}{2}\|x_{t,k}-x_{t,k}^{\text{ref}}\|_{W_x}^2
        + \tfrac{1}{2}\|u_{t,k}-u_{t,k}^{\text{ref}}\|_{W_u}^2\\
        \text{s.t.} \quad &\quad x_{t,0}=\hat{x}_t,\\
        &\quad x_{t,k+1}=F(x_{t,k},u_{t,k}), ~k\in\mathbb{Z}_0^{N-1},\\
        &\quad \underline{u}\leq u_{t,k}\leq \Bar{u}, \quad\quad \quad \quad  k\in\mathbb{Z}_0^{N-1},
    \end{aligned}
\end{equation}
where $N$ is the prediction horizon length; $\hat{x}_t$ is the feedback state at time $t$; $x_{t,k}\in\mathbb{R}^{n_x}$ and $u_{t,k}\in\mathbb{R}^{n_u}$ are the $k$th state and control input along the prediction horizon length at time $t$, respectively;  $\mathbf{x}_t^{\text{ref}}$ and $\mathbf{u}_t^{\text{ref}}$ are the given reference trajectories at time $t$; and $F(\cdot)$ denotes the discrete-time nonlinear dynamics, which is obtained from the continuous-time nonlinear dynamics via the Runge-Kutta 4 (RK4) method. The constraints $[\underline{u},\Bar{u}]$ come from the physical limitations on the control inputs (e.g., actuators).

\begin{figure*} 
\begin{equation}\label{eqn_QP_NMPC}
\begin{aligned}
&\text{QP}_{\text{NMPC}}(\hat{x}_t,\mathbf{x}_t^{\text{guess}},\mathbf{u}_t^{\text{guess}},\mathbf{x}_t^{\text{ref}},\mathbf{u}_t^{\text{ref}})\triangleq\argmin\tfrac{1}{2}\|\Delta x_{t,N}\|_{W_N}^2 + \Delta x_{t,N}^\top W_N^\top(x_{t,N}^{\text{guess}}-x_{t,N}^{\text{ref}})+ \tfrac{1}{2}\sum_{k=0}^{N-1}\|\Delta x_{t,k}\|_{W_x}^2\\
    & \qquad\qquad\qquad\qquad\qquad\qquad\quad\qquad\qquad\qquad  + \Delta x_{t,k}^\top W_x^\top(x_{t,k}^{\text{guess}}-x_{t,k}^{\text{ref}})  + \tfrac{1}{2}\|\Delta u_{t,k}\|_{W_u}^2 + \Delta u_{t,k}^\top W_u^\top(u_{t,k}^{\text{guess}}-u_{t,k}^{\text{ref}})\\
    &\text{s.t.}\quad\Delta x_{t,0}=\hat{x}_t-x_{t,0}^{\text{guess}}\\
           &\, \qquad\Delta x_{t,k+1}=A_{t,k}\Delta x_{t,k} + B_{t,k}\Delta u_{t,k}+r_{t,k}, \quad k\in\mathbb{Z}_0^{N-1},\\
           &\,\qquad\underline{u}-u_{t,k}^{\text{guess}}\leq\Delta u_{t,k}\leq\Bar{u}-u_{t,k}^{\text{guess}},\qquad\quad\quad\quad \,k\in\mathbb{Z}_0^{N-1}.
        \end{aligned}
    \end{equation}  
\end{figure*}
    
Apart from using sequential quadratic programming methods to solve $\text{NLP}(\hat{x}_t,\mathbf{x}_t^{\text{ref}},\mathbf{u}_t^{\text{ref}})$, a well-known and successful technique is the RTI scheme. This scheme reduces the computational time of an NMPC problem to that of a linear MPC problem, while providing nominal stability  \cite{diehl2005nominal}. The key ingredients of RTI-based NMPC are
\begin{itemize}
    \item[i)] \textit{Single Full Newton Step:} 
    
    At the time $t$, assuming that a good initial guess $(\mathbf{x}_t^{\text{guess}},\mathbf{u}_t^{\text{guess}})$ is given, the full Newton step 
    \[
(\mathbf{x}_t,\mathbf{u}_t)\leftarrow(\mathbf{x}_t^{\text{guess}},\mathbf{u}_t^{\text{guess}}) + (\Delta\mathbf{x}_t,\Delta\mathbf{u}_t)
    \]
    provides an excellent approximation of the fully converged NMPC solution, 
    where $(\Delta\mathbf{x}_t,\Delta\mathbf{u}_t)$ is the solution of the QP problem $\text{QP}_{\text{NMPC}}\equiv \text{QP}_{\text{NMPC}}(\hat{x}_t,\mathbf{x}_t^{\text{guess}},\mathbf{u}_t^{\text{guess}},\mathbf{x}_t^{\text{ref}},\mathbf{u}_t^{\text{ref}})$ in \eqref{eqn_QP_NMPC}.

    Here the trajectories $\Delta\mathbf{x}_t=(\Delta x_{t,0},\cdots{},\Delta x_{t,N})$ and $\Delta\mathbf{u}_t=(\Delta u_{t,0},\cdots{},\Delta u_{t,N})$ are the deviation between the system trajectories $\mathbf{x}_t,\mathbf{u}_t$ and the guess trajectories $\mathbf{x}_t^{\text{guess}}$, $\mathbf{u}_t^{\text{guess}}$,
    \[
     \Delta x_{t,k} = x_{t,k} - x_{t,k}^{\text{guess}}, \Delta u_{t,k} = u_{t,k} - u_{t,k}^{\text{guess}}, k\in\mathbb{Z}_0^{N-1}
    \]
    and $A_{t,k},B_{t,k},r_{t,k}$ stem from linearizing the discrete-time nonlinear dynamic $F$ at time $t$,
\begin{equation}
\label{eqn_A_B_r_discrete_time}
        \begin{aligned}
           & A_{t, k}=\left.\frac{\partial F(x, u)}{\partial x}\right|_{x_{t, k}^{\text{guess}}, u_{t, k}^{\text{guess}}},&k\in\mathbb{Z}_0^{N-1},\\
           & B_{t, k}=\left.\frac{\partial F(x, u)}{\partial u}\right|_{x_{t, k}^{\text{guess}}, u_{t, k}^{\text{guess}}},&k\in\mathbb{Z}_0^{N-1},\\
           & r_{t,k} = F(x_{t,k}^{\text{guess}},u_{t,k}^{\text{guess}}) - x_{t,k+1}^{\text{guess}},&k\in\mathbb{Z}_0^{N-1}.
        \end{aligned} 
    \end{equation}
    \item[ii)] \textit{Shifting Initialization based on Previous Solution:} 
    
    A very good initial guess at time $t$ can be constructed from the shifting of a good solution obtained at the previous time $t-1$. Let $(\mathbf{x}_{t-1},\mathbf{u}_{t-1})$ be the solution of $\text{QP}_{\text{NMPC}}$ at time $t-1$, then the initial guess $(\mathbf{x}_{t}^{\text{guess}},\mathbf{u}_{t}^{\text{guess}})$ is constructed from
\begin{equation}
\label{eqn_shift}
        \begin{aligned}
        x_{t,k}^{\text{guess}} &= x_{t-1,k+1}, ~\quad k\in\mathbb{Z}_0^{N-1}, \\
        u_{t,k}^{\text{guess}} &= u_{t-1,k+1}, \quad\  k\in\mathbb{Z}_0^{N-2},\\
        u_{t,N-1}^{\text{guess}} &= u_{t,N-2}^{\text{guess}} = u_{t-1,N-1}, \\
        x_{t,N}^{\text{guess}} &= f(x_{t,N-1}^{\text{guess}},u_{t,N-1}^{\text{guess}}).
    \end{aligned}
    \end{equation}
    Note that shifting the given reference trajectories $(\mathbf{x}_t^{\text{ref}},\mathbf{u}_t^{\text{ref}})$ to construct the initial guess could lead to poor performance when the actual system trajectory is not in the neighborhood of the reference.
    \item[iii)] \textit{Preparation-Feedback Split:} 
    
    The RTI-based NMPC and the linear MPC both only require solving one QP problem on-line, but their difference is that the QP of linear MPC is constructed off-line once whereas the resulting QP of RTI-based NMPC is constructed on-line. To further reduce the computational delay, RTI-based NMPC splits the on-line computation into two phases, 
    \begin{itemize}
        \item \textit{A preparation phase}, completing the part of computations related to the $\text{QP}_{\text{NMPC}}$ construction as much as possible before the arrival of the feedback state $\hat{x}_t$;
        \item \textit{A feedback phase}, finishing the QP construction and solving the $\text{QP}_{\text{NMPC}}$ upon the feedback state $\hat{x}_t$ arriving.
    \end{itemize}
    In this manner, the computational time of RTI-based NMPC is only marginally larger than linear MPC. 
\end{itemize}

\subsection{Computing the sensitivities $\{A_{t,k},B_{t,k},r_{t,k}\}_{k=0}^{N-1}$ from continuous-time systems}
In the context of NMPC, the nonlinear model is usually derived from physical laws as a continuous-time, autonomous nonlinear differential equation,
\begin{equation}
    \dot{x}(t)=f(x(t),u(t)).
\end{equation}
In this case, computing the sensitivities $\{A_{t,k},B_{t,k},r_{t,k}\}_{k=0}^{N-1}$ from (\ref{eqn_A_B_r_discrete_time}) is not straightforward. For example,  the discrete-time nonlinear dynamic $F$ is obtained from continuous-time $f$ via the RK4 method to preserve both integration accuracy and computation efficiency. Here, we illustrate how to compute the sensitivities $\{A_{t,k},B_{t,k},r_{t,k}\}_{k=0}^{N-1}$ without resorting to defining $F$ explicitly. Denote 
\[
\begin{aligned}
    f_x(x(t),u(t))&=\left.\frac{\partial f(x,u)}{\partial x}\right|_{x=x(t),u=u(t)},\\
    f_u(x(t),u(t))&=\left.\frac{\partial f(x,u)}{\partial u}\right|_{x=x(t),u=u(t)},
\end{aligned}
\]
and consider the $N_s$ integration steps over the sampling time $\Delta t$ resulting in $t_i=\frac{\Delta t}{N_s}$ discretization time, Procedure \ref{procedure_computing_A_B_r} shows how to compute the $k$th sensitivities $A_{t,k},B_{t,k},r_{t,k}$ from the continuous-time dynamic $f$ via the RK4 method. Procedure \ref{procedure_computing_A_B_r} needs to be executed $N$ times to calculate all $\{A_{t,k},B_{t,k},r_{t,k}\}_{k=0}^{N-1}$ in the \textit{preparation phase}.
\begin{lemma}\label{lemma_flop_sensitivities}
    Let $m_f, m_{f_x},m_{f_u}$ denote the flops required by the evaluation of $f(\cdot),f_x(\cdot),f_u(\cdot)$, respectively. Then the flops required by the computation of the sensitivities $\{A_{t,k},B_{t,k},r_{t,k}\}_{k=0}^{N-1}$ is 
    \[
    \begin{aligned}
    & N\times(n_x+n_u+ N_s(4m_f+4m_{f_x}+4m_{f_u}+8n_x^3+8n_x^2n_u\\
    &\qquad\quad\ \ +10n_x^2+10n_xn_u+16n_x)+n_x^2+n_xn_u+n_x).
    \end{aligned}
    \]
\end{lemma}

\floatname{algorithm}{Procedure}
\begin{algorithm}[t]
    \caption{Computing $A_{t,k},B_{t,k},r_{t,k}$}\label{procedure_computing_A_B_r}
    \textbf{Input}: $x_{t,k}^{\text{guess}},u_{t,k}^{\text{guess}},x_{t,k+1}^{\text{guess}}$, the number of discretization steps $N_s$, the discretization time $t_i=\frac{\Delta t}{N_s}$, $f_x(\cdot)$ and $f_u(\cdot)$.
    \vspace*{.1cm}\hrule\vspace*{.1cm}
    \begin{enumerate}[label*=\arabic*., ref=\theenumi{}]
        \item $x=x_{t,k}^{\text{guess}}, u=u_{t,k}^{\text{guess}},A=I,B=0$
        \item \textbf{for} $j=1,\ldots, N_s$ \textbf{do}
        \begin{enumerate}[label=\theenumi{}.\arabic*., ref=\theenumi{}.\arabic*]
            \item $\kappa(1)\leftarrow f(x,u)$
            \item $[\kappa_x(1),\kappa_u(1)]\leftarrow f_x(x,u)[A,B]+[0,f_u(x,u)]$
            \item $\kappa(2)\leftarrow f(x+\tfrac{1}{2}t_i\kappa(1),u)$
            \item $[\kappa_x(2),\kappa_u(2)]\leftarrow f_x(x+\tfrac{1}{2}t_i\kappa(1),u)[A+\tfrac{1}{2}t_i \kappa_x(1)$, $B + \tfrac{1}{2} t_i\kappa_u(1)] + [0,f_u(x+\tfrac{1}{2}t_i\kappa(1),u)]$
            \item $\kappa(3)\leftarrow f(x+\tfrac{1}{2}t_i\kappa(2),u)$
            \item $[\kappa_x(3),\kappa_u(3)]\leftarrow f_x(x+\tfrac{1}{2}t_i\kappa(2),u)[A+\tfrac{1}{2}t_i \kappa_x(2)$, $B+\tfrac{1}{2}t_i\kappa_u(2)] + [0,f_u(x+\tfrac{1}{2}t_i\kappa(2),u)]$
            \item $\kappa(4)\leftarrow f(x+t_i\kappa(3),u)$
            \item $[\kappa_x(4),\kappa_u(4)]\leftarrow f_x(x+t_i\kappa(3),u)[A+t_i\kappa_x(3),$ $B+t_i\kappa_u(3)] + [0,f_u(x+t_i\kappa(3),u)]$
            \item $x\leftarrow x+ \frac{t_i}{6}(\kappa(1)+2\kappa(2)+2\kappa(3)+\kappa(4))$
            \item $[A,B]\leftarrow [A,B] + \frac{t_i}{6}([\kappa_x(1),\kappa_u(1)]+2[\kappa_x(2),\kappa_u(2)]$ $+\,2[\kappa_x(3),\kappa_u(3)]+[\kappa_x(4),\kappa_u(4)])$
        \end{enumerate}
    \item \textbf{end}.
    \item $A_{t,k}\leftarrow A, B_{t,k}\leftarrow B, r_{t,k}\leftarrow x-x_{t,k+1}^{\text{guess}}$
    \end{enumerate}
    \vspace*{.1cm}\hrule\vspace*{.1cm}
    \textbf{Output}: $A_{t,k},B_{t,k},r_{t,k}$.
\end{algorithm}

\subsection{$\text{QP}_{\text{NMPC}}$ construction}
By using a condensing construction that eliminates the states, $\text{QP}_{\text{NMPC}}$ (\ref{eqn_QP_NMPC}) can be formulated as a box-constrained QP (Box-QP). Our time-certified IPM algorithm (see Section \ref{sec_time_certified_IPM}) is tailored for the Box-QP with the unit box constraint $[-e,e]$. After generating $\{A_{t,k},B_{t,k},r_{t,k}\}_{k=0}^{N-1}$, we first scale the control inputs $\Delta u_{t,k}$ subject to $[-e,e]$, denoted as $\Delta \bar{u}_{t,k}$, through
\begin{equation}
    \Delta u_{t,k} = D\Delta \Bar{u}_{t,k} + d_{t,k},
\end{equation}
where 
\begin{equation}\label{eqn_D_d}
\begin{aligned}
D&=\tfrac{1}{2}\diag(\Bar{u}-\underline{u}),\\
d_{t,k}&=\tfrac{1}{2}(\Bar{u}+\underline{u})-u_{t,k}^{\text{guess}},\quad k\in\mathbb{Z}_0^{N-1}.
\end{aligned}    
\end{equation}
Then, the associated terms are updated as
\begin{equation}\label{eqn_B_r}
\begin{aligned}
    \Bar{B}_{t,k}&= B_{t,k}D,&k\in\mathbb{Z}_0^{N-1},\\
    \Bar{r}_{t,k}&= r_{t,k} + B_{t,k}d_{t,k},&k\in\mathbb{Z}_0^{N-1}.\\
\end{aligned}    
\end{equation}
Define $z\triangleq \operatorname{col}(\Delta\bar{u}_{t,0},\cdots{},\Delta\bar{u}_{t,N-1})\in\mathbb{R}^{n}$, 
where $n=N n_u$, $\bar{Q}\triangleq\diag(W_x,\cdots{},W_x,W_N)$, $\bar{R}\triangleq\diag(DW_uD,\cdots{},DW_uD)$, and
\begin{footnotesize}
\begin{equation}\label{eqn_S}
S\triangleq\left[\begin{array}{cccc}
      \bar{B}_{t,0} &  0 & \cdots & 0\\
      A_{t,0}\bar{B}_{t,0} & \bar{B}_{t,1} & \cdots & 0\\
      \vdots & \vdots & \ddots & \vdots\\
      \prod\limits_{k=0}^{N-1}A_{t,k}\bar{B}_{t,0} & \prod\limits_{k=0}^{N-2}A_{t,k}\bar{B}_{t,1} & \cdots & \bar{B}_{t,N-1}
\end{array}\right].    
\end{equation}
\end{footnotesize}
Then construct $\text{QP}_{\text{NMPC}}$ (\ref{eqn_QP_NMPC}) as the scaled Box-QP
\begin{subequations}\label{problem_Box_QP}
    \begin{align} 
        z^*=&\argmin_z J(z)= \tfrac{1}{2}z^\top H z + z^\top h\label{problem_Box_QP_objective}\\
        &\,\textrm{s.t.}\ -e \leq z \leq e,\label{proble_Box_QP_box_constraint}
    \end{align}
\end{subequations}
where the Hessian matrix is given by
\begin{equation}\label{eqn_H}
    H=\bar{R}+S^{\top}\bar{Q}S,
\end{equation}
and the gradient vector is given by
\begin{equation}\label{eqn_h}
    h = S^{\top}\bar{Q}g+\! \left[\begin{array}{c}
        DW_u(\frac{1}{2}(\bar{u}+\underline{u})-u_{t,0}^{\text{ref}}) \\
        \vdots\\
       DW_u(\frac{1}{2}(\bar{u}+\underline{u})-u_{t,N-1}^{\text{ref}})
    \end{array}\right]\!,
\end{equation}
in which the vector $g=g_1+g_2$ and
\begin{subequations}\label{eqn_g}
    \begin{align}
    g_1&\triangleq\mathbf{x}_t^{\text{guess}} -\mathbf{x}_t^{\text{ref}}
    +\!\left[\begin{array}{c}
    \bar{r}_{t,0} \\
    A_{t,1}\bar{r}_{t,0}+\bar{r}_{t,1}\\
    \vdots\\
    \sum\limits_{k=1}^{N-1}A_{t,k}\bar{r}_{t,0}+\cdots+\bar{r}_{t,N-1}
 \end{array}\right]\label{eqn_g1}\!,\\
    g_2&\triangleq\!\left[\begin{array}{c}
    A_{t,0} \\
    A_{t,1}A_{t,0} \\
    \vdots \\
    \prod\limits_{k=0}^{N-1}A_{t,k}
    \end{array}\right]\!(\hat{x}_t-x_{t,0}^{\text{guess}})\label{eqn_g2}    
    \end{align}
\end{subequations}

\begin{lemma}\label{lemma_H_h_computation}
    The computation of $\{\bar{B}_{t,k},\bar{r}_{t,k}\}_{k=0}^{N-1}$ requires $N(3n_xn_u+n_x)$ flops, the computation of $S$ requires $(N^2-N)n_xn_u^2$ flops, the computation of $H$ requires $(2N^3+N^2+N)n_x^2n_u+Nn_u^2$ flops, the computation of $g_1$ requires $2Nn_x+2(N-1)n_x^2$ flops, the computation of $g_2$ requires $n_x+2Nn_x^2$ flops, and the computation of $h$ requires $2Nn_x^2+(N^2+N)n_xn_u+(N^2-N)n_x+n_u+N(2n_u+n_u^2)$ flops.
\end{lemma}

\begin{remark}\label{remark_H_avoids}
By Lemma \ref{lemma_H_h_computation}, the computation of $H$ dominates, especially when $N$ is considerably larger than $n_x$ and $n_u$. Consequently, circumventing the computation of $H$ has the potential to diminish the computational time of the \textit{preparation phase}. The subsequent section demonstrates that the Riccati recursion technique not only eliminates the need to compute $H$ but also lowers the cost associated with the Newton step.
\end{remark}

\section{Time-certified Riccati-based IPM Algorithm}\label{sec_time_certified_IPM}
Based on the path-following full-Newton IPM, our recent work \cite{wu2023direct} proposed an \textit{direct} optimization algorithm to solve the Box-QP (\ref{problem_Box_QP}). Its Karush–Kuhn–Tucker (KKT) conditions are 
\begin{subequations}\label{eqn_KKT}
\begin{align}
    Hz + h + \gamma - \theta = 0\label{eqn_KKT_a},\\
    z + \alpha - e=0\label{eqn_KKT_b},\\
    z - \omega + e=0\label{eqn_KKT_c},\\
    \gamma \phi = 0\label{eqn_KKT_d},\\
    \theta \psi = 0\label{eqn_KKT_e},\\
    (\gamma,\theta,\alpha,\omega)\geq0.
\end{align}
\end{subequations}
A positive parameter $\tau$ was introduced by the path-following IPM to replace (\ref{eqn_KKT_d}) and (\ref{eqn_KKT_e}) by
\begin{subequations}\label{eqn_KKT_tau}
\begin{align}
    \gamma \phi = \tau^2 e\label{eqn_KKT_tau_d},\\
    \theta \psi = \tau^2 e\label{eqn_KKT_tau_e}.
\end{align}
\end{subequations}
As $\tau$ tends to 0, the path $(z_{\tau},\gamma_{\tau},\theta_{\tau},\phi_{\tau},\psi_{\tau})$ converges to a solution of \eqref{eqn_KKT}. The feasible variants of the path-following IPM algorithm boast the best theoretical $O(\sqrt{n})$ iteration complexity \cite{wright1997primal}. Therefore, our algorithm is based on feasible IPM in which all iterates are required to be strictly in the feasible set
\[
\mathcal{F}^0\triangleq\big\{(z,\gamma,\theta,\phi,\psi)\big\lvert~\eqref{eqn_KKT_a}\textrm{--}\eqref{eqn_KKT_c}\text{ satisfied},(\gamma,\theta,\phi,\psi)>0\big\}.
\]

\subsection{Strictly feasible initial point}
In our recent work \cite{wu2023direct}, a cost-free initialization strategy was proposed to find a strictly feasible initial point that also satisfies the specific conditions. A strictly feasible initial point is
\[
z^0 = 0,\gamma^0=\|h\|_\infty - \tfrac{1}{2}h,\theta^0 =\|h\|_\infty + \tfrac{1}{2}h, \phi^0 = e, \\\psi^0 = e,
\]
where $\|h\|_\infty=\max \{ |h_1|,|h_2|,\cdots{},|h_{n}|\}$. It is straightforward to show that the above initial point strictly lies in $\mathcal{F}^0$.

\begin{remark}[Initialization strategy]\label{remark_initialization_strategy}
For $h=0$, the optimal solution of problem (\ref{problem_Box_QP}) is $z^*=0$; for $h\neq0$, first scale the objective (\ref{problem_Box_QP_objective}) (which does not change the optimal solution) as
\[
\min_z \tfrac{1}{2} z^\top \!\left(\frac{2\lambda}{\|h\|_\infty}H\right) \!z + z^\top \!\left(\frac{2\lambda}{\|h\|_\infty}h\right)\!.
\]
With the definitions $\tilde{H} = \frac{1}{\|h\|_\infty}H$ and $\tilde{h}=\frac{1}{\|h\|_\infty}h$, $\|\tilde{h}\|_\infty=1$ and (\ref{eqn_KKT_a}) can be replaced by
\[
2\lambda \tilde{H}z+2\lambda\tilde{h}+\gamma-\theta=0,
\]
and the initial points
\begin{equation}\label{eqn_initialization_stragegy}
z^0 = 0,\ \gamma^0= 1 - \lambda \tilde{h},\ \theta^0 =1 + \lambda \tilde{h},\ \phi^0 = e,\  \psi^0 = e
\end{equation}
\end{remark}
can be adopted, where
$\lambda =\tfrac{1}{\sqrt{n+1}}$. 
It is straightforward to verify that (\ref{eqn_initialization_stragegy}) lies in $\mathcal{F}^0$. The reason to use the scale factor $\frac{2\lambda}{\|h\|_\infty}$ is to make the initial point satisfy the neighborhood requirements, e.g., see \cite[Lemma 4]{wu2023direct}.

\subsection{Newton direction}
Denote $v=\operatorname{col}(\gamma, \theta) \in \mathbb{R}^{2n}$ and $s=\operatorname{col}(\phi,\psi) \in \mathbb{R}^{2n}$. Then replace (\ref{eqn_KKT_tau_d}) and (\ref{eqn_KKT_tau_e}) by $v s = \tau^2e$ to obtain the new complementary condition,
\begin{equation}
    \sqrt{vs} = \sqrt{\tau^2e}\label{eqn_new_complementary}.
\end{equation}
From \textit{Remark \ref{remark_initialization_strategy}}, $(z, v, s)\in \mathcal{F}^0$ and a direction $(\Delta z,\Delta v,\Delta s)$ can be obtained by solving the system of linear equations,
\begin{subequations}\label{eqn_newKKT_compact}
    \begin{align}
        2\lambda\tilde{H}\Delta z + \Omega \Delta v =& 0\label{eqn_newKKT_compact_a},\\
        \Omega^\top \Delta z + \Delta s = & 0\label{eqn_newKKT_compact_b},\\
        \sqrt{\frac{s}{v}}\Delta v + \sqrt{\frac{v}{s}}\Delta s = &2(\tau e-\sqrt{v s}),\label{eqn_newKKT_compact_c}
    \end{align}
\end{subequations}
where $\Omega=[I,-I] \in\mathbb{R}^{n \times 2n}$. Letting
\begin{subequations}\label{eqn_Delta_gamma_theta_phi_psi}
    \begin{align}
        &\Delta \gamma=\frac{\gamma}{\phi}\Delta z+2\!\left(\sqrt{\frac{\gamma}{\phi}}\tau e-\gamma\right)\!,\\
        &\Delta \theta=-\frac{\theta}{\psi}\Delta z+2\!\left(\sqrt{\frac{\theta}{\psi}}\tau e-\theta\right)\!,\\
        &\Delta\phi = - \Delta z,\\
        &\Delta\psi = \Delta z
    \end{align}
\end{subequations}
reduces (\ref{eqn_newKKT_compact}) into a more compact system of linear equations,
\begin{equation}{\label{eqn_compact_linsys}}
    \begin{aligned}
&\left(2\lambda\tilde{H}+\diag\!\left(\frac{\gamma}{\phi}\right)\! + \diag\!\left(\frac{\theta}{\psi}\!\right)\! \right) \!\Delta z\\
&\qquad\qquad\qquad\qquad=2\!\left(\sqrt{\frac{\theta}{\psi}}\tau e-\sqrt{\frac{\gamma}{\phi}}\tau e+ \gamma - \theta\right)
    \end{aligned}
\end{equation}

\subsection{Implementation of the Newton step via factorized Riccati recursion}
Solving the linear system (\ref{eqn_compact_linsys}) typically requires $O(N^3n_u^3)$ flops. Specifically, when employing the Cholesky decomposition method, the required flops are $\frac{1}{6}N^3n_u^3+\frac{5}{2}N^2n_u^2+\frac{1}{3}Nn_u$. In general, the RTI-based NMPC scheme demonstrates effective performance provided that the sampling time $\Delta t$ is short and the time prediction horizon $T_p$ is long \cite{gros2020linear}. This often leads to a prolonged prediction horizon length $N=\frac{T_p}{\Delta t}$, e.g., of $20,40,$ and $60$ in practical applications. 

This article adopts the Riccati recursion to solve the linear system (\ref{eqn_compact_linsys}) efficiently when the ratio $\frac{n_x}{n_u}$ is not large and the prediction horizon $N$ is long. Subsequently, we illustrate how the Riccati recursion not only minimizes the flops needed to solve (\ref{eqn_compact_linsys}) but also circumvents the computation of $H$.

The linear system (\ref{eqn_compact_linsys}) is equivalent to the formulation
\begin{equation}\label{problem_lin_solve}
\begin{aligned}
     \min &\ \ \frac{1}{2}\Delta z^{\top}\!\left(2\lambda\tilde{H}+\diag\!\left(\frac{\gamma}{\phi}\right)\!+\diag\!\left(\frac{\theta}{\psi}\right)\!\right)\!\Delta z \\
     &\qquad\qquad- 2\Delta z^{\top} \!\left(\sqrt{\frac{\theta}{\psi}}\tau e-\sqrt{\frac{\gamma}{\phi}}\tau e+ \gamma - \theta\right).
\end{aligned}
\end{equation}
Let $\mathrm{col}(\Delta \hat{u}_0,\cdots{},\Delta \hat{u}_{N-1})\triangleq\Delta z$, $\mathrm{col}(\Delta\hat{x}_1,\cdots{},\Delta\hat{x}_N)\triangleq S\Delta z$, $\Delta\hat{x}_0=0$, and $\hat{Q}_{k+1}\triangleq\frac{2\lambda}{\|h\|_\infty}W_x,k\in\mathbb{Z}_{0}^{N-2}$, $\hat{Q}_{N}\triangleq\frac{2\lambda}{\|h\|_\infty}W_N$, for $k\in\mathbb{Z}_{0}^{N-1}$. Exploiting the structure of $S$ in (\ref{eqn_S}) and $H$ in (\ref{eqn_H}), gives that
\[
\begin{aligned}
&\hat{R}_k\triangleq\frac{2\lambda}{\|h\|_\infty}DW_uD+\diag\!\left(\frac{\gamma}{\phi}+\frac{\theta}{\psi}\right)_{\!kn_u+1:(k+1)n_u},\\
&\hat{g}_k=-2\left(\sqrt{\frac{\theta}{\psi}}\tau e-\sqrt{\frac{\gamma}{\phi}}\tau e+ \gamma - \theta\right)_{\!kn_u+1:(k+1)n_u}.
\end{aligned}
\] 
The above formulation (\ref{problem_lin_solve}) is equivalent to the unconstrained linear quadratic regulator problem,
\begin{equation}\label{problem_LQR}
    \begin{aligned}
       \min\sum_{k=0}^{N-1} &\left(\tfrac{1}{2}\|\Delta\hat{u}_k\|^2_{\hat{R}_k}+\hat{g}_k^{\top}\Delta\hat{u}_t+\tfrac{1}{2}\|\Delta\hat{x}_{k+1}\|^{2}_{\hat{Q}_{k+1}} \right)\\
     \textrm{s.t.}\quad  & \Delta\hat{x}_0=0,\\
       \quad &\Delta\hat{x}_{k+1}=A_{t,k}\Delta\hat{x}_k+\bar{B}_{t,k}\Delta\hat{u}_k,\quad  k\in\mathbb{Z}_{0}^{N-1},\\
    \end{aligned}
\end{equation}
which can be solved efficiently by Riccati recursion. Instead of adopting the classical Riccati recursion, this article adopts the factorized Riccati recursion to further reduce the computational cost \cite{frison2013efficient}, e.g., as shown in Procedure \ref{procedure_implementation_Newton_step_factorized}.
\floatname{algorithm}{Procedure}
\begin{algorithm}
    \caption{Implementation of the Newton step via factorized Riccati recursion}\label{procedure_implementation_Newton_step_factorized}
    \textbf{Input}: $\{A_{t,k},\bar{B}_{t,k},\hat{Q}_k,\hat{R}_k,\hat{g}_k\}_{k=0}^{N-1},\hat{Q}_N$
    \vspace*{.1cm}\hrule\vspace*{.1cm}
    $L_N\leftarrow \mathrm{chol}(\hat{Q}_N)$\\
    $p_N\leftarrow 0$\\
    \textbf{for} $k=N-1,\cdots{},0$ \textbf{do}
    \begin{enumerate}[label*=\arabic*., ref=\theenumi{}]
        \item[] $[L_k^{\top}\bar{B}_{t,k},L_k^{\top}A_{t,k}]\leftarrow L_{k+1}^{\top}\cdot_{\mathrm{dtrmm}}\![\bar{B}_{t,k},A_{t,k}]$
        \item[] $\mathcal{O}\leftarrow [L_k^{\top}\bar{B}_{t,k},L_t^{\top}A_{t,k}]^{\top} \cdot_{\mathrm{dsyrk}}\![L_t^{\top}\bar{B}_{t,k},L_k^{\top}A_{t,k}]$
        \item[] $\footnotesize\left[\begin{array}{@{}cc@{}}
            \Lambda_k &  \\
            M_k & L_k
        \end{array}\right]\!\leftarrow \mathrm{chol}_{L}\biggl(\mathcal{O}+\biggl[\begin{array}{@{}cc@{}}
            \hat{R}_k &  \\
             & \hat{Q}_{k}
        \end{array}\biggr] \biggr)$
        \item[] $q_k\leftarrow(\Lambda_{k}\Lambda_{k}^{\top})^{-1}(\bar{B}_{t,k}^{\top}p_{k+1}+\hat{g}_k)$
        \item[] $p_k\leftarrow  A_{t,k}^{\top}p_{k+1} - M_k\Lambda_k^{\top}q_k $
    \end{enumerate}
    \textbf{end}\\
    $\Delta\hat{x}_0\leftarrow 0$\\
    \textbf{for} $k=0,\cdots{},N-1$ \textbf{do}
    \begin{enumerate}[label*=\arabic*., ref=\theenumi{}]
    \item[] $\Delta\hat{u}_k\leftarrow -\Lambda_k^{-\top}M_k^{\top}\Delta\hat{x}_k-q_k$ 
    \item[] $\Delta\hat{x}_{k+1}\leftarrow A_{t,k}\Delta\bar{x}_k+\bar{B}_{t,k}\Delta\hat{u}_k$
    \end{enumerate}
    \textbf{end}
    \vspace*{.1cm}\hrule\vspace*{.1cm}
    \textbf{Output}: $\Delta z\leftarrow \mathrm{col}(\Delta \bar{u}_0,\cdots{},\Delta \bar{u}_{N-1})$
\end{algorithm}
\begin{lemma}[See \cite{frison2013efficient}]\label{lemma_flop_Riccati}
    Procedure \ref{procedure_implementation_Newton_step_factorized} requires a total of
    \[
    N(\tfrac{7}{3}n_x^3+4n_x^2n_u+2n_xn_u^2+\tfrac{1}{3}n_u^3) + N(8n_x^2+8n_xn_u+2n_u^2)~\text{[flops]}.
    \] 
\end{lemma}

\subsection{Iteration complexity and algorithm implementation}
The complete RTI-based input-constrained MPC scheme is summarized as Algorithm \ref{alg_time_certifed_IPM}. 
Next, we derive the iteration complexity. Denote $\beta\triangleq \sqrt{vs}$ and define the proximity measure as
\begin{equation}\label{eqn_xi}
\xi(\beta,\tau)=\tfrac{\|\tau e-\beta\|}{\tau}.
\end{equation}
\begin{lemma}[See \cite{wu2023direct}]\label{lemma_strictly_feasible}
Let $\xi:=\xi(\beta,\tau) < 1$. Then the full Newton step is strictly feasible, i.e., $v_{+}>0$ and $s_{+}>0$.
\end{lemma}
\begin{lemma}[See \cite{wu2023direct}]\label{lemma_duality_gap}
    After a full Newton step, let $v_{+}=v+\Delta v$ and $s_{+}=s+\Delta s$; then the duality gap is
    \[
        v_{+}^{\top}s_{+}\leq(2n)\tau^2.
    \]
\end{lemma}
\begin{lemma}[See \cite{wu2023direct}]\label{lemma_xi}
Suppose that $\xi=\xi(\beta,\tau)<1$ and $\tau_+=(1-\eta)\tau$ where $0<\eta<1$. Then,
\[
\xi_+=\xi(\beta_+,\tau_+)\leq\tfrac{\xi^2}{1+\sqrt{1-\xi^2}}+\tfrac{\eta\sqrt{2n}}{1-\eta}.
\]
Furthermore, if $\xi\leq\frac{1}{\sqrt{2}}$ and $\eta=\frac{\sqrt{2}-1}{\sqrt{2n}+\sqrt{2}-1}$, then $\xi_+\leq\frac{1}{\sqrt{2}}$.
\end{lemma}
\begin{lemma}[See \cite{wu2023direct}]\label{lemma_xi_condition}
The value of $\xi(\beta,\tau)$ before the first iteration is denoted as
$\xi^0=\xi(\beta^0,(1-\eta)\tau^0)$. If $(1-\eta)\tau^0=1$ and $\lambda=\frac{1}{\sqrt{n+1}}$, then $\xi^0\leq\frac{1}{\sqrt{2}}$ and $\xi(\beta, w)\leq\frac{1}{\sqrt{2}}$ are always satisfied.
\end{lemma}
\begin{lemma}[See \cite{wu2023direct}]\label{lemma_exact}
    Let $\eta=\frac{\sqrt{2}-1}{\sqrt{2n}+\sqrt{2}-1}$ and $\tau^0=\frac{1}{1-\eta}$, Algorithm \ref{alg_time_certifed_IPM} exactly requires    \begin{equation}
\mathcal{N}=\!\left\lceil\frac{\log\frac{2n}{\epsilon}}{-2\log\frac{\sqrt{2n}}{\sqrt{2n}+\sqrt{2}-1}}\right\rceil\! + 1
    \end{equation}  
    iterations, the resulting vectors being $v^\top s\leq\epsilon$. 
\end{lemma}

Lemmas \ref{lemma_flop_sensitivities}, \ref{lemma_H_h_computation}, \ref{lemma_flop_Riccati}, \ref{lemma_exact} imply the totals in Theorem 1.
\begin{theorem}\label{theorem_1}
    In the \textit{preparation phase} of Algorithm \ref{alg_time_certifed_IPM}, Step 1 takes $Nn_x+Nn_u+m_f$ [flops], Step 2 takes $N(n_x+n_u+N_s(4m_f+4m_{f_x}+4m_{f_u}+8n_x^3+8n_x^2n_u+10n_x^2+10n_xn_u+16n_x)+n_x^2+n_xn_u+n_x)$ [flops], Step 3 takes $n_u+Nn_u+N(2n_xn_u+n_x)+(N^2-N)n_xn_u^2+2Nn_x+2(N-1)n_x^2$ [flops];  In the \textit{feedback phase} of Algorithm \ref{alg_time_certifed_IPM}, Step 1 takes $n_x+4Nn_x^2+(N^2+N)n_xn_u+(N^2-N)n_x+n_u+N(2n_u+n_u^2)$ [flops], Step 2 takes  $Nn_u$ [flops], Step 3 takes $5Nn_u+3$ [flops], Step 4 takes $\mathcal{N}(1+N(\tfrac{7}{3}n_x^3+4n_x^2n_u+2n_xn_u^2+\tfrac{1}{3}n_u^3)+ N(8n_x^2+8n_xn_u+2n_u^2)+15Nn_u +5n_x)$ [flops], Step 5 takes $n_x$ [flops], Step 6 takes $N(2n_u+n_x^2+n_xn_u+2n_x)$ [flops], Step 7 takes $(N+1)n_x+Nn_u$[flops].
\end{theorem}

\floatname{algorithm}{Algorithm}
\begin{algorithm}
    \caption{A time-certified IPM algorithm for RTI-based input-constrained MPC (\ref{eqn_QP_NMPC}) at sampling time $t$
    }\label{alg_time_certifed_IPM}
    \textit{Preparation phase: performed over the time interval $[t-\Delta t,t]$}\\
    \textbf{Input}: previous solution $(\mathbf{x}_{t-1},\mathbf{u}_{t-1})$, reference $(\mathbf{x}_t^{\text{ref}},\mathbf{u}_t^{\text{ref}})$, continuous-time dynamic $f(\cdot)$, and its partials $f_x(\cdot), f_u(\cdot)$;
    \begin{enumerate}[label*=\arabic*., ref=\theenumi{}]
        \item Construct $(\mathbf{x}_{t}^{\text{guess}},\mathbf{u}_{t}^{\text{guess}})$ according to (\ref{eqn_shift}) from the previous solution $(\mathbf{x}_{t-1},\mathbf{u}_{t-1})$;
        \item Calculate $\{A_{t,k},B_{t,k},r_{t,k}\}_{k=0}^{N-1}$ by performing Procedure \ref{procedure_computing_A_B_r} $\forall k\in\mathbb{Z}_0^{N-1}$;
        \item Calculate the scaling factors $D,\{d_{t,k}\}_{k=0}^{N-1}$ from (\ref{eqn_D_d}), $\{\bar{B}_{t,k},\bar{r}_{t,k}\}_{k=0}^{N-1}$ from (\ref{eqn_B_r}), $S$ from (\ref{eqn_S}), and $g_1$ from (\ref{eqn_g1})
    \end{enumerate}
    \textbf{return} $D,\{d_{t,k}\}_{k=0}^{N-1}$, $\{A_{t,k},\bar{B}_{t,k},r_{t,k}\}_{k=0}^{N-1}$ and $S, g_1$
    \vspace*{.1cm}\hrule\vspace*{.1cm}
    \textit{Feedback phase: performed at time t upon arrival of $\hat{x}_t$}\\
    \textbf{Input}: the feedback states $\hat{x}_t$, $D$, $\{d_{t,k}\}_{k=0}^{N-1}$, $\{A_{t,k},\bar{B}_{t,k}$ $,\bar{r}_{t,k}\}_{k=0}^{N-1}$, $S$, $g_1$, the stopping tolerance $\epsilon$; the required exact number of iterations $\mathcal{N}=\left\lceil\frac{\log\frac{2n}{\epsilon}}{-2\log\frac{\sqrt{2n}}{\sqrt{2n}+\sqrt{2}-1}}\right\rceil\!+1$.
    \begin{enumerate}[label*=\arabic*., ref=\theenumi{}]
        \item Calculate $g_2$ from (\ref{eqn_g2}) and $h$ from (\ref{eqn_h}) based on $\hat{x}_t, S, g_1$;
        \item \textbf{if }$\|h\|_\infty=0$, $z\leftarrow0$ and go to Step 5; \textbf{otherwise},
        \item Initialize $(z,\gamma,\theta,\phi,\psi)$ from (\ref{eqn_initialization_stragegy}) where $\lambda\leftarrow\frac{1}{\sqrt{n+1}}$, $\eta\leftarrow\frac{\sqrt{2}-1}{\sqrt{2n}+\sqrt{2}-1}$ and $\tau\leftarrow\frac{1}{1-\eta}$;
        \item \textbf{for} $i=1, 2,\cdots{}, \mathcal{N}$ \textbf{do}
        \begin{enumerate}[label=\theenumi{}.\arabic*., ref=\theenumi{}.\arabic*]
        \item $\tau\leftarrow(1-\eta)\tau$;
        \item Obtain $\Delta z$ by performing Procedure \ref{procedure_implementation_Newton_step_factorized};
        \item Calculate $(\Delta\gamma,\Delta\theta,\Delta\alpha,\Delta\omega)$ from \eqref{eqn_Delta_gamma_theta_phi_psi};
        \item  $z\leftarrow z+\Delta z$, $\gamma\leftarrow \gamma+\Delta \gamma$, $\theta\leftarrow \theta+\Delta \theta$, $\alpha\leftarrow \alpha+\Delta \alpha$, $\omega\leftarrow \omega+\Delta \omega$;
        \end{enumerate}
        \item[] \textbf{end}
        \item $\Delta x_{t,0}\leftarrow\hat{x}_t-x_{t,0}^{\text{guess}}$;
        \item for $k=0,1,\cdots{},N-1$
        \[
        \begin{aligned}
            \Delta u_{t,k}&\leftarrow D z_{kn_u+1:kn_u+n_u}+d_{t,k}\\
            \Delta x_{t,k+1}&\leftarrow A_{t,k}\Delta x_{t,k}+\bar{B}_{t,k} z_{kn_u+1:kn_u+n_u} + \bar{r}_{t,k}
        \end{aligned}
        \]
        \item Apply the full Newton step 
        \[
(\mathbf{x}_{t},\mathbf{u}_t)\leftarrow(\mathbf{x}_{t}^{\text{guess}},\mathbf{u}_t^{\text{guess}}) + (\Delta\mathbf{x}_{t},\Delta\mathbf{u}_t)
        \]
    \end{enumerate}
    \textbf{return} NMPC solution $(\mathbf{x}_{t},\mathbf{u}_t)$
\end{algorithm}

\section{Numerical Example: Chaotic Lorenz Stabilization}
The Lorenz system is a well-known nonlinear dynamical system to model convective hydrodynamic flows \cite{lorenz1963deterministic}. The consequence of chaos is that, under certain parameterizations, the trajectories of the deterministic Lorenz system become arbitrarily sensitive to small perturbations in the initial conditions. Slightly different trajectories separate vastly over long horizons, which is popularly coined as the \textit{butterfly effect}.

In this work, we address the stabilization task of applying affine control signals to each coordinate of the Lorenz system,
\begin{equation}
    \begin{aligned}
        \dot{x} &= \sigma(y-x) + u_x,\\
        \dot{y} &= x(\rho-z)-y+u_y,\\
        \dot{z} &= xy - \beta z + u_z,
    \end{aligned}
    \label{eqn_lorenz}
\end{equation}
where $(x,y,z)$ are the states, and $(u_x,u_y,u_z)\in[-3,3]^3$ are the control inputs. The parameters $\sigma=10,~\rho=28,~\beta=8/3$ are chosen so that \ref{eqn_lorenz} exhibits chaotic behavior with two strange attractors, $(\pm \sqrt{\beta(\rho-1)}, \pm \sqrt{\beta(\rho-1)}, \rho-1)$. Stabilizing the chaotic system to one of its attractors $(6\sqrt{2},6\sqrt{2},27)$ is a nontrivial task due to its fractal-like trajectories and heightened sensitivity to small perturbations. Here, the desired reference is $x_{t,k}^{\text{ref}}=\mathrm{col}(6\sqrt{2},6\sqrt{2},27),~u_{t,k}^{\text{ref}}=\mathrm{col}(0,0,0), k\in\mathbb{Z}_0^{N-1}$. The weight matrices were chosen as $W_N=W_x=I$ and $W_u=0.1I$ for the states and control inputs, respectively. The sampling time is $\Delta t=0.01$ s, and the time prediction horizon is $T_p=0.2$ s, so the prediction horizon length is $N=\frac{T_p}{\Delta t}=20$. Additionally, we opt for $N_s=2$ integration steps to calculate the sensitivities.
\begin{figure*}[!t]\label{fig}
\begin{picture}(140,110)
\put(-15,-10){\includegraphics[width=75mm]{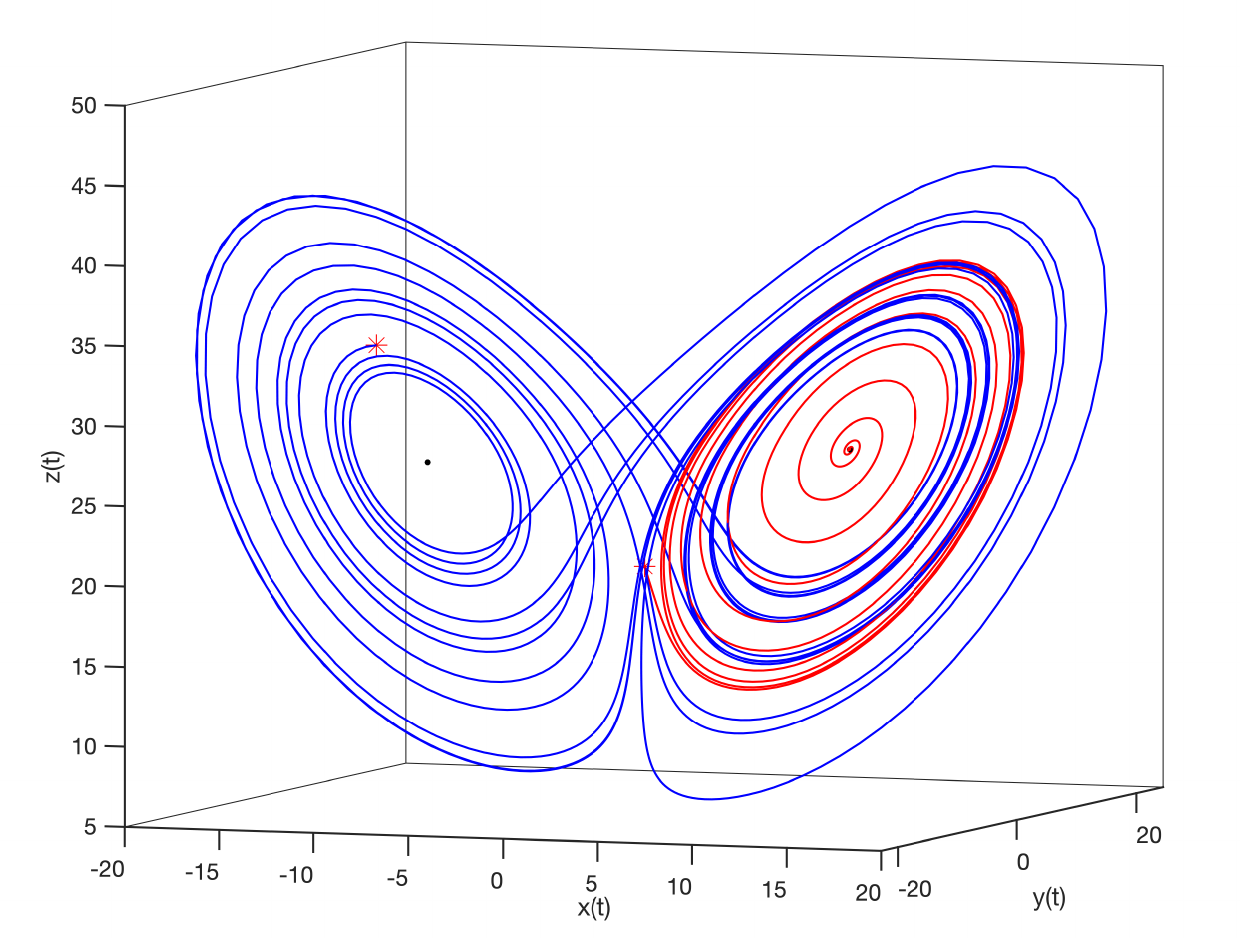}}
\put(190,0){\includegraphics[width=55mm]{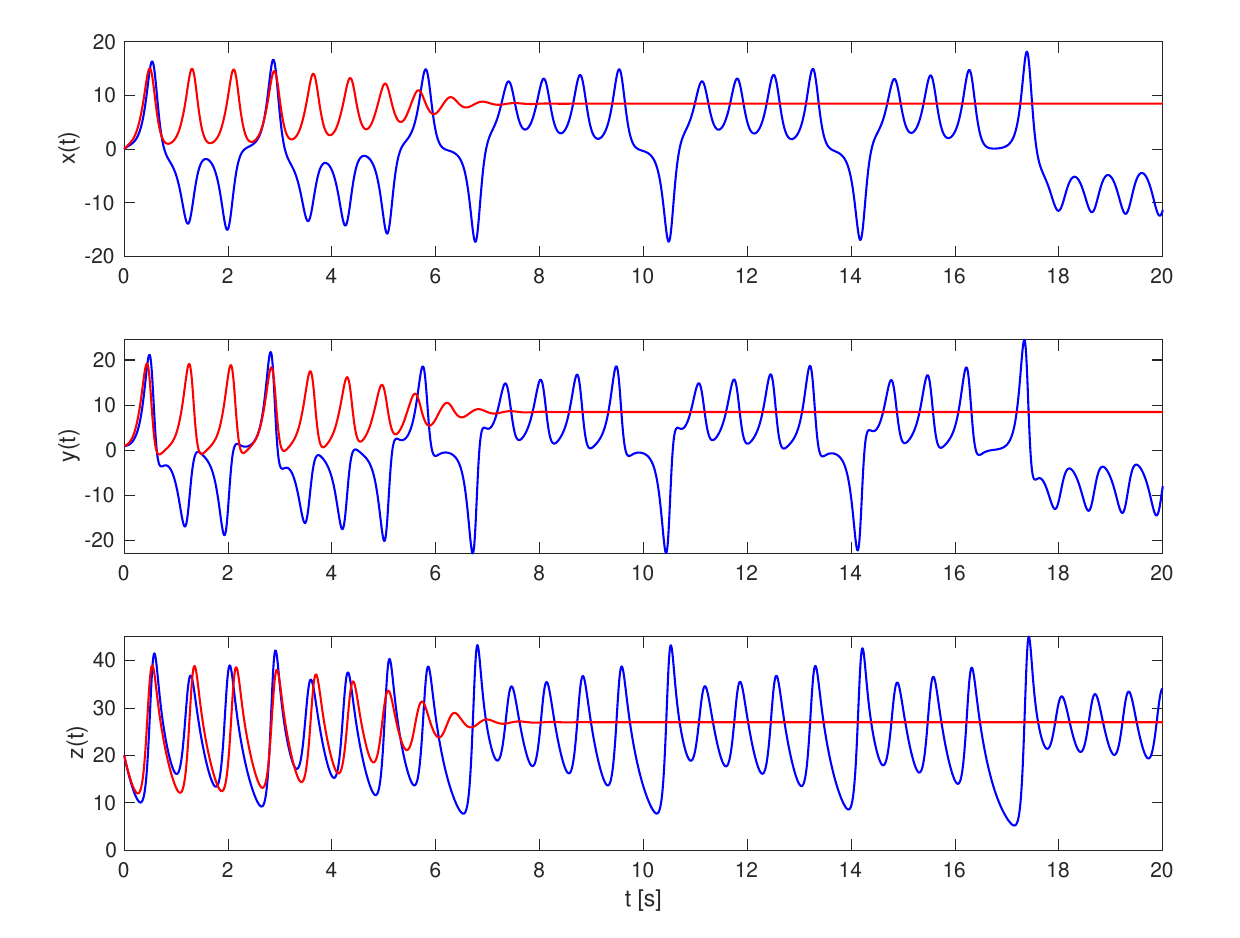}}
\put(340,0){\includegraphics[width=55mm]{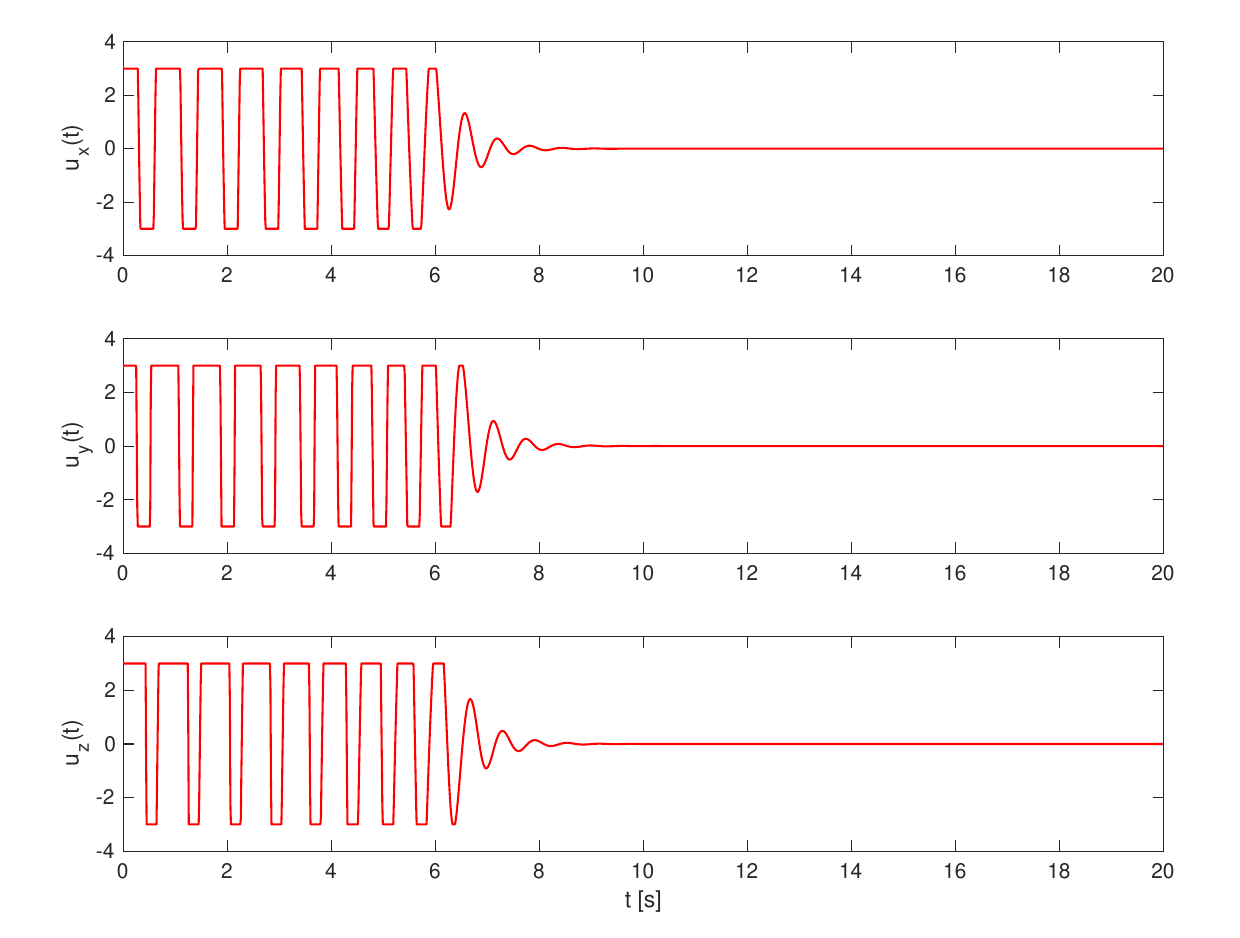}}
\end{picture}
\caption{Closed-loop simulation of the chaotic Lorenz system with the RTI-based input-constrained MPC. The blue and red lines represent the results of uncontrolled and MPC controlled, respectively. Left: phase trajectory. Middle: time evolution of the three coordinates. Right: the three control inputs.}
\label{fig_Lorenz}
\end{figure*}

Before doing the closed-loop simulation, we can exactly calculate the flops required by the \textit{preparation phase} and \textit{feedback phase} of Algorithm \ref{alg_time_certifed_IPM} at each sampling time. The dimension of the Box-QP (\ref{problem_Box_QP}) is $n=3$$\times$$N=3$$\times$$20=60$, and we adopt the stopping criteria $\epsilon=10^{-6}$, so Algorithm \ref{alg_time_certifed_IPM} will exactly perform $\mathcal{N}=\!\left\lceil\frac{\log\frac{2\times 60}{10^{-6}}}{-2\log\frac{\sqrt{2\times 60}}{\sqrt{2\times 60}+\sqrt{2}-1}}\right\rceil\! + 1= 252$
iterations in the \textit{feedback phase}. The Lorenz system flux expressions $f(\cdot)$, $f_x(\cdot)$, and $f_u(\cdot)$ take $m_f=10, ~m_{f_x}=4,$ and $m_{f_u}=0$ (since in $f_x(\cdot)$ only 4 elements vary and $f_u(\cdot)=[1,1,1]^\top$) [flops], respectively. Thus, by Theorem \ref{theorem_1}, the \textit{preparation phase} requires a total of $4.05$$\times$$10^4$ [flops] and the \textit{feedback phase} requires a total of $2.23$$\times$$10^6$ flops. Altogether, the algorithm execution requires a wall time of $0.0025$ s on a personal laptop with 1 Gflop/s computing power (i.e., a trivial requirement for most processors today). Hence, we can obtain a certificate ensuring that the execution time will be less than the specified sampling time of $\Delta t=0.01$ s.

Algorithm~\ref{alg_time_certifed_IPM} is executed in MATLAB2023a via a C-mex interface, and the closed-loop simulation was performed on a contemporary MacBook Pro with 2.7~GHz 4-core Intel Core i7 processors and 16GB RAM. The number of iterations is exactly $252$, and the execution time of the two phases adds up to about $0.002$ s which is less than $\Delta t=0.01$ s. The closed-loop simulation results are depicted in Fig.~\ref{fig_Lorenz}, which illustrates that the MPC effectively stabilizes the chaotic Lorenz system to the specified attractor, in stark contrast to the \textit{butterfly effect} trajectories of the uncontrolled case. The three control inputs do not violate $[-3,3]$ and eventually converge to zeros.

\section{Conclusion}
This article proposes an execution-time-certified algorithm for RTI-based input-constrained NMPC problems. To address the MPC setting with a long prediction horizon that RTI-based NMPC often results in, the factorized Riccati recursion, whose computation cost scales linearly to the prediction horizon, is used to efficiently solve the Newton system at each iteration. 
In the future, we will continue work on extensions to general execution-time-certified NMPC algorithms including both input and state constraints.

\section{Acknowledgement}
This work was supported by the U.S. Food and Drug Administration under the FDA BAA-22-00123 program, Award Number 75F40122C00200. Krystian Ganko was also supported by the U.S. Department of Energy, Office of Science, Office of Advanced Scientific Computing Research, Department of Energy Computational Science Graduate Fellowship under Award Number DE-SC0022158.

\bibliographystyle{IEEEtran}
\bibliography{ref} 
\end{document}